\documentclass[
  reprint,
  superscriptaddress,
  amsmath,amssymb,
  aps,
  prl,
  longbibliography
]{revtex4-2}

\usepackage{graphicx}
\usepackage{dcolumn}
\usepackage{bm}
\usepackage{siunitx}
\usepackage{makecell}
\usepackage[colorlinks=true, allcolors=blue]{hyperref}
\usepackage{placeins}

\begin{document}

\title{Nanoscale Dipolar Fields in Artificial Spin Ice Probed by Scanning NV Magnetometry}

\author{Ephraim Spindler}
\email[E-Mail: ]{spindler@rptu.de}
\affiliation{RPTU University Kaiserslautern-Landau, Department of Physics and State Research Center OPTIMAS, Kaiserslautern, Germany}

\author{Vinayak Shantaram Bhat}
\altaffiliation{Present address: Thomas J. Watson College of Engineering and Applied Science, Binghamton University, Binghamton, NY 13902, USA}
\affiliation{Department of Physics \& Astronomy, University of Delaware, Newark, DE 19716, USA}

\author{Elke Neu}
\affiliation{RPTU University Kaiserslautern-Landau, Department of Physics and State Research Center OPTIMAS, Kaiserslautern, Germany}

\author{Mathias Weiler}
\affiliation{RPTU University Kaiserslautern-Landau, Department of Physics and State Research Center OPTIMAS, Kaiserslautern, Germany}

\author{M. Benjamin Jungfleisch}
\affiliation{Department of Physics \& Astronomy, University of Delaware, Newark, DE 19716, USA}

\begin{abstract}
We investigate dipolar coupling fields in two square-lattice artificial spin ice (ASI) systems with different lattice constants using scanning probe microscopy based on a single nitrogen-vacancy (NV) center in diamond. This technique offers nanoscale spatial resolution, operates under ambient conditions, and provides quantitative stray field measurements, making it uniquely suited for studying nanoscale magnetic textures. Our approach combines fluorescence quenching imaging and continuous-wave optically detected magnetic resonance (cODMR). A comparison of the two ASI samples, which differ in their lattice constants of \qty{1000}{\nano\meter} and \qty{910}{\nano\meter} respectively, reveals differences in the appearance of ice-rule violations - deviations from the lowest energy configuration in ASI vertices. We attribute these variations to varying coupling strengths dictated by the lattice constant. From the cODMR data, we extract both axial and transverse components of the local magnetic field relative to the NV axis. Micromagnetic modeling of these measurements allows for an iterative determination of the external magnetic field orientation, the detection of subtle magnetization tilts induced by weak external fields (well below the nanomagnets' switching threshold), and an estimation of the effective saturation magnetization, thereby accounting for deviations in nanomagnet dimensions. These findings provide crucial insights into the tunable magnetic interactions in ASI, paving the way for the design of advanced magnonic and spintronic devices.
\end{abstract}

\maketitle

\begin{figure*}[htb]
\centering
\includegraphics[width= \textwidth]{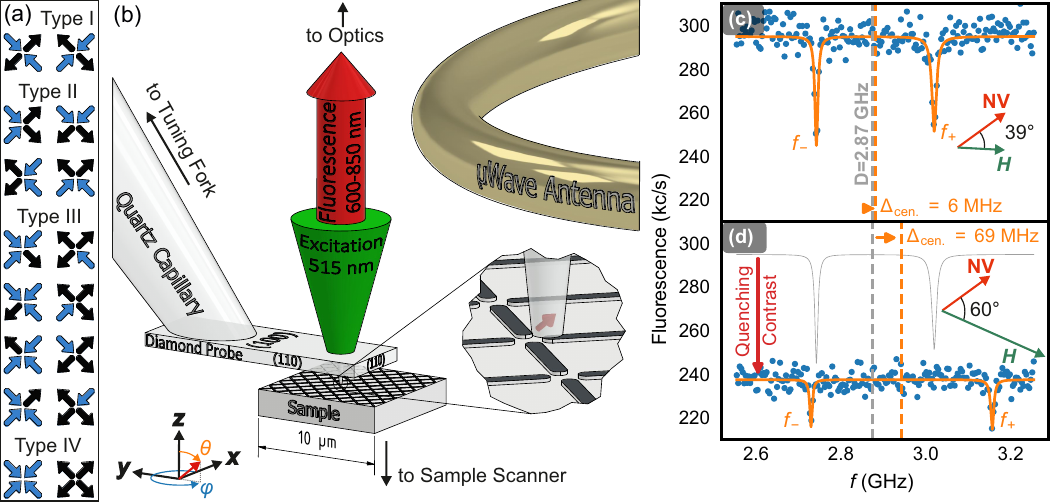}
\caption{
(a) Overview of the 16 possible moment configurations, divided into 4 topological vertex types.\cite{wang_artificial_2006}
(b) Schematic of the scanning NV magnetometry setup including the ASI sample \textit{S-A}.
(c) cODMR spectrum recorded at a magnetic field of $|\mu_0\boldsymbol{H}| = \qty{6.3}{\milli\tesla}$, oriented at \ang{39} relative to the NV axis. The experimental data (blue) is fitted with a double Lorentzian (orange), yielding ESR frequencies $f_-$ and $f_+$. The orange dashed line indicates the center frequency $(f_+ + f_-)/2$, whose deviation from the zero-field splitting $D$ (gray dashed line) is denoted as $\Delta_\mathrm{cen}$.
(d) cODMR spectrum recorded at $|\mu_0\boldsymbol{H}| = \qty{15.1}{\milli\tesla}$ and a relative angle of \ang{60}. Acquisition parameters are identical to (c), except for the magnetic field. The fit from (c) is overlaid in gray for comparison. An arrow indicates the reduction of the off-resonant fluorescence rate by about \qty{20}{\percent} compared to (c).}
\label{fig:Setup}
\end{figure*}

Artificial spin ice (ASI) systems \cite{Sultana_2025,skjaervo_advances_2019,wang_artificial_2006, KAFFASH2021127364}, composed of interacting nanomagnets arranged in periodic or aperiodic lattices, offer a versatile platform for exploring phenomena such as magnetic monopoles \cite{ladak_direct_2010}, collective dynamics \cite{Montoncello_2018,KAFFASH2021127364}, and topological effects \cite{Drisko2017}. The 16 possible vertex configurations in square ASI geometries are depicted in Fig.~\ref{fig:Setup}(a), divided into 4 topological vertex types. In such systems, frustration manifests through the ice rule \cite{harris_geometrical_1997}, which dictates a two-in/two-out magnetization configuration at each vertex (type I and II). Violations of this rule (type III and IV) lead to monopole-like excitations \cite{castelnovo_magnetic_2008, ladak_direct_2010} and a rich spectrum of emergent behaviors \cite{skjaervo_advances_2019,Lendinez_2021}. 

Probing local magnetic states and interaction fields in ASIs at the nanoscale remains a significant challenge. While techniques like X-ray magnetic circular dichroism photoemission electron microscopy (XMCD-PEEM) \cite{element_specific, le_guyader_studying_2012, farhan_direct_2013, acosta_temperature_2010} and magnetic force microscopy (MFM) \cite{wang_artificial_2006,Gartside_2018} provide high resolution, they each have drawbacks. XMCD-PEEM requires synchrotron access and ultra-high vacuum, whereas an MFM's magnetic tip can perturb the sample, and the extended interaction volume of the tip complicates the extraction of local field information. Scanning magnetometry using single nitrogen-vacancy (NV) centers in diamond \cite{rondin_magnetometry_2014} overcomes many of these limitations, offering non-invasive, quantitative field mapping with nanoscale resolution under ambient conditions. 

Since the theoretical proposal of a scanning single-spin microscope \cite{chernobrod_spin_2005}, its specific conceptualization using NV centers \cite{degen_scanning_2008} and its first experimental realization \cite{balasubramanian_nanoscale_2008} as a scanning probe technique, NV magnetometry has matured into a robust platform \cite{maletinsky_robust_2012} for investigating diverse condensed matter systems. Its effectiveness has been demonstrated through the imaging of current densities in nanoscale conductor networks \cite{chang_nanoscale_2017}, the visualization of complex non-collinear antiferromagnetic orders \cite{gross_real-space_2017}, and the quantitative probing of magnetism in two-dimensional van der Waals crystals \cite{thiel_probing_2019}. These applications highlight the technique's unique ability to provide calibrated, vectorial dipolar field information in environments where maintaining the sample's magnetic integrity is crucial.

However, applying scanning NV magnetometry to ASI systems is not trivial.
The strong surface magnetic fields of strongly interacting nanomagnet lattices can cause significant fluorescence quenching and loss of ODMR contrast due to the mixing of NV spin states in the presence of large transverse field components \cite{rondin_magnetometry_2014}. 
To mitigate these effects we employ a lifted probe approach, performing measurements at a controlled distance from the sample surface. Furthermore, the application of a bias magnetic field must be carefully optimized: it must be strong enough to ensure clear separation of the NV resonance frequencies, and include a perpendicular component to enhance off-axis sensitivity, while remaining below the threshold where excessive field-induced quenching occurs.
In this work, we develop a systematic methodology to quantitatively assess dipolar stray fields in artificial spin ice by combining scanning single-NV magnetometry with micromagnetic modeling.
The developed framework is readily adaptable to a wide range of magnetic nanostructures and complex textures, providing a general pathway for the quantitative study of local interaction fields in functional magnetic materials.

We investigate two square-lattice ASI samples with varied lattice constants but identical nanomagnet dimensions to enable a controlled comparison of dipolar coupling strengths. We combine fluorescence quenching imaging and continuous-wave optically detected magnetic resonance (cODMR) to identify vertex configurations and extract vector-resolved magnetic field components. By comparing these measurements with micromagnetic simulations, we detect subtle magnetization tilts induced by weak external magnetic fields and infer the effective saturation magnetization through an iterative process that matches the simulations with the experimental results.

The ASI arrays are fabricated using a \qty{100}{\kilo\electronvolt} electron beam lithography system. Following the development of the mask, a \qty{25}{\nano\meter}-thick Ni\textsubscript{81}Fe\textsubscript{19} film, capped with a \qty{3}{\nano\meter} Al layer, is deposited via electron beam evaporation. Each nanomagnet measures approximately \qty{110}{\nano\meter} in width and \qty{810}{\nano\meter} in length. The lattice constants for samples \textit{S-A} and \textit{S-B} are \qty{1000}{\nano\meter} and \qty{910}{\nano\meter}, respectively. To facilitate full saturation under a magnetic field applied along the x-axis, all nanomagnets are rotated by \qty{45}{\degree}. Further details on sample fabrication and characterization are provided in Section S1 of the supplementary material (SM).

The experimental setup of the scanning single-NV magnetometer (ProteusQ, Qnami AG) is depicted in Fig.~\ref{fig:Setup}(b). The sample is raster-scanned beneath a stationary diamond probe (MX series, (100)-oriented with sides along the (110) orientation), which houses an NV center at the apex of a nanopillar on the diamond's lower surface. The NV depth is expected to be $9\pm4~\mathrm{nm}$, based on the nitrogen ion implantation energy of \qty{6}{\kilo\electronvolt}. The NV axis orientation is first described by the polar and azimuthal angles $\theta = \qty{54.75}{\degree}$ and $\varphi = \qty{270}{\degree}$ in the laboratory frame. This corresponds to one of the four possible $\varphi$ angles (\qty{0}{\degree}, \qty{90}{\degree}, \qty{180}{\degree}, and \qty{270}{\degree}) dictated by the $\langle 111 \rangle$ crystallographic axes of the diamond sensor. We choose this orientation based on the NV response to a known magnetic field (see Section S2 of the SM for details). 

It is important to acknowledge that a deviation from these nominal angles by a few degrees is very common due to misalignments during probe fabrication, sensor attachment to the quartz capillary, or probe mounting in the microscope. In our analysis, we account for this by performing an in-situ calibration and refinement of the NV angles during the final evaluation stage. This step can be skipped if the orientation is already calibrated beforehand—for example, using a rotating magnetic field or an approach based on scanning across perpendicularly magnetized microdisks \cite{khera2026methods}.

The probe is affixed to a quartz capillary mounted on a tuning fork for frequency-modulated shear-force feedback. A \qty{515}{\nano\meter} laser beam is focused onto the nanopillar via a microscope objective ($\mathrm{NA} = 0.7$), which simultaneously collects the red fluorescence emitted by the NV center. This fluorescence is spectrally filtered using a long pass filter with a cut-on wavelength of \qty{650}{\nano\meter} and detected using a single-photon counting module. A wire loop-shaped microwave antenna  positioned close to the diamond probe (see Fig.~\ref{fig:Setup}(b)) drives the NV electron spin resonance (ESR), while a permanent magnet assembly positioned below the sample provides a tunable magnetic field.

We define the magnetic field components along and perpendicular to the NV axis as the axial field $H_{\text{NV}\parallel}$ and the off-axis field $H_{\text{NV}\perp}$, respectively. These components are derived from the resonance frequencies $f_-$ and $f_+$ using the relations provided by Van Der Sar et al. \cite{van_der_sar_nanometre-scale_2015}:
\begin{widetext}
\begin{align}
\mu_0 H_{\text{NV}\parallel} &= \frac{ 
  \sqrt{
    -\left(D + f_{+} - 2f_{-}\right)
    \left(D + f_{-} - 2f_{+}\right)
    \left(D + f_{+} + f_{-}\right)
  }
}{3 \gamma \sqrt{3D}} \label{eq:B_para}, \\
\mu_0 H_{\text{NV}\perp} &= \frac{
  \sqrt{
    -\left(2D - f_{+} - f_{-}\right)
    \left(2D - f_{+} + 2f_{-}\right)
    \left(2D - f_{-} + 2f_{+}\right)
  }
}{3 \gamma \sqrt{3D}}\label{eq:B_perp}.
\end{align}
\end{widetext}
Here, $D = \qty{2.87}{\giga\hertz}$ \cite{scanning_1997} represents the zero-field splitting in the electronic ground state, $\gamma = \qty{28}{\giga\hertz \per \tesla}$ is the electron gyromagnetic ratio, and $\mu_0$ is the vacuum permeability.
In the small-field approximation, the Zeeman shift of the NV center's ESR frequencies is linear and symmetric to $D$, depending only on the parallel magnetic field component, $H_{\text{NV}\parallel}$. However, Eqs.~\eqref{eq:B_para} and \eqref{eq:B_perp} describe the more general case for a non-negligible perpendicular field component. While the dominant response to $H_{\text{NV}\parallel}$ remains linear and symmetric, $H_{\text{NV}\perp}$ induces an assymmetric frequency shift with a quadratic dependency, shifting both transitions to higher frequencies \cite{beaver_optimizing_2024, welter_scanning_2022}.

Assuming Lorentzian line shapes, the magnetic sensitivities to axial and off-axis fields are calculated according to Dréau et al. \cite{dreau_avoiding_2011} and Baever et al. \cite{beaver_optimizing_2024} as:
\begin{align}
   \mu_0\eta_{\parallel} &= \frac{4}{3 \sqrt{3}} \frac{\Delta \nu}{C \sqrt{R}} \frac{1}{\gamma} \label{eq:sen_para}\\
   \mu_0\eta_{\perp} &= \frac{4}{3 \sqrt{3}} \frac{\Delta \nu}{C \sqrt{R}} \frac{D}{3 \gamma^2 \mu_0 H_{\text{NV}\perp}}\label{eq:sen_perp}
\end{align} 
Here, $\Delta \nu$ is the full width at half maximum (FWHM) ODMR linewidth, $R$ the fluorescence rate without microwave excitation of the ESR and $C = \frac{R - R_\mathrm{res}}{R}$ the ODMR contrast with the on-resonance fluorescence rate $R_\mathrm{res}$.

Figure~\ref{fig:Setup}(c,d) presents representative cODMR spectra corresponding to relatively low and high values of $H_{\text{NV}\perp}$, respectively, acquired during a scan of sample \textit{S-A} (see Fig.~\ref{fig:details}(c)). The difference in orientation and strength of $\mu_0\boldsymbol{H}$ is caused by the local ASI stray field adding up with the homogeneous external field. Comparing these spectra visually demonstrates the magnetic contrast mechanisms employed in this study, namely the linear Zeeman shift originating from $H_{\text{NV}\parallel}$, and both the non-linear Zeeman shift and fluorescence quenching caused by $H_{\text{NV}\perp}$. The dashed gray line marks the zero-field splitting $D$, while the dashed orange line indicates the center frequency $(f_+ + f_-)/2$. Its deviation from $D$ highlights the non-linear Zeeman shift.

Table~\ref{tab:contrast} lists the parameters extracted from the Lorentzian fit, as well as the resulting sensitivities derived from Eqs.~\eqref{eq:sen_para} and \eqref{eq:sen_perp} for both cases. It shows that a moderate off-axis bias field enhances the off-axis magnetic sensitivity $\eta_{\perp}$, consistent with reports by Beaver et al. \cite{beaver_optimizing_2024}. Moreover, such a bias field helps suppress potential errors stemming from thermal drift of $D$.

\begin{table}[htbp]
\caption{\label{tab:contrast}%
Parameters $\Delta\nu$, $C$ and $R$ extracted from the fits shown in Fig.~\ref{fig:Setup}(c,d) and averaged over both resonances, $\mu_0H_{\text{NV}\perp}$ computed from $f_-$ and $f_+$ using Eq. \eqref{eq:B_perp} and sensitivities $\mu_0\eta_{\parallel}$ and $\mu_0\eta_{\perp}$ computed using Eqs.~\eqref{eq:sen_para} and \eqref{eq:sen_perp}.}
\begin{ruledtabular}
\begin{tabular}{lcccccr}
\makecell{Panel\footnote{corresponding to the cODMR spectra shown in Fig.~\ref{fig:Setup}(c,d)}} &
\makecell{$\Delta\nu$ \\ (MHz)} &
\makecell{$C$} &
\makecell{$R$ \\ (kc/s)} &
\makecell{$\mu_0H_{\text{NV}\perp}$ \\ (mT)} &
\makecell{$\mu_0\eta_{\parallel}$ \\ ($\mu$T/$\sqrt{\text{Hz}}$)} &
\makecell{$\mu_0\eta_{\perp}$ \\ ($\mu$T/$\sqrt{\text{Hz}}$)} \\
\colrule
(c) & 10.5 & 0.16 & 295 & 4.0 & 3.4 & 28.9\\
(d) & 7.7 & 0.09 & 237 & 13.0 & 4.6 & 12.2\\
\end{tabular}
\end{ruledtabular}
\end{table}

A secondary effect of strong off-axis fields is fluorescence quenching due to spin state mixing \cite{tetienne_magnetic-field-dependent_2012}, observable in the cODMR spectra as a \qty{20}{\percent} reduction in off-resonant fluorescence (indicated by a red arrow in Fig.~\ref{fig:Setup}(d)). Although this effect reduces ODMR contrast, fluorescence quenching offers a qualitative, but fast and all-optical method for probing sufficiently strong (typically $> \qty{5}{\milli\tesla}$) off-axis fields. Typical integration times for quenching images range from \qty{10}{\milli\second} to \qty{30}{\milli\second} per pixel.

\begin{figure*}[htbp!]
    \centering
    \includegraphics{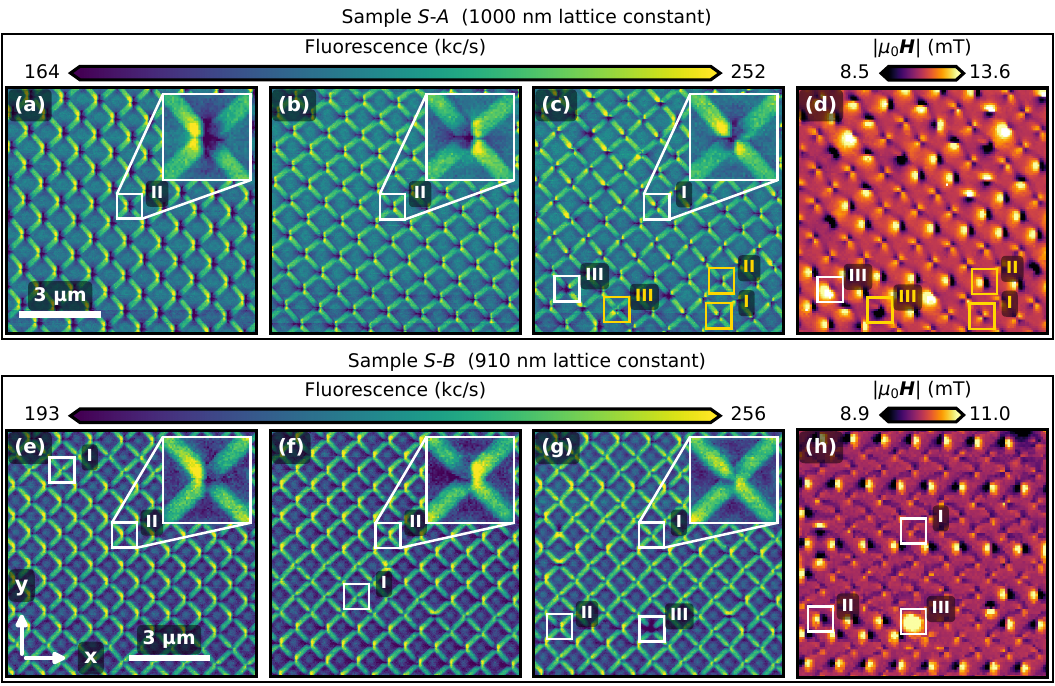}
\caption{
    Overview of magnetic states in samples \textit{S-A} (first row) and \textit{S-B} (second row). Fluorescence quenching images (a)–(c) and (e)–(g) were acquired in contact mode, while lifted-probe cODMR scans (d) and (h) were measured with a lift height of \qty{100}{\nano\meter}. All panels show the same regions of samples \textit{S-A} and \textit{S-B}, respectively, after application of external magnetic fields (applied along the x-direction, following saturation in the opposite direction):
    (a) \qty{-35.8}{\milli\tesla},
    (b) \qty{-40.3}{\milli\tesla},
    (c), (d) \qty{-38.3}{\milli\tesla} (for \textit{S-A}),
    (e) \qty{-33.9}{\milli\tesla},
    (f) \qty{-38.5}{\milli\tesla}, and
    (g), (h) \qty{-36.3}{\milli\tesla} (for \textit{S-B}).
    Insets in panels (a)–(c) and (e)–(g) display representative vertices, highlighted by frames labeled with their corresponding vertex types. In panels (e) and (f), additional type~I vertices are highlighted to emphasize that the observed state is not a fully saturated type~II lattice, unlike in panels (a) and (b).
    For panels (c) and (d), as well as (g) and (h), type~I, II, and III vertices are consistently labeled, and the same vertices are framed to facilitate direct comparison between the two imaging modes.
    Three specific vertices (highlighted in gold) are selected for in-depth analysis in Fig.~\ref{fig:details}.
}\label{fig:overview}
\end{figure*}

\begin{figure*}[htbp!]
    \centering
    \includegraphics{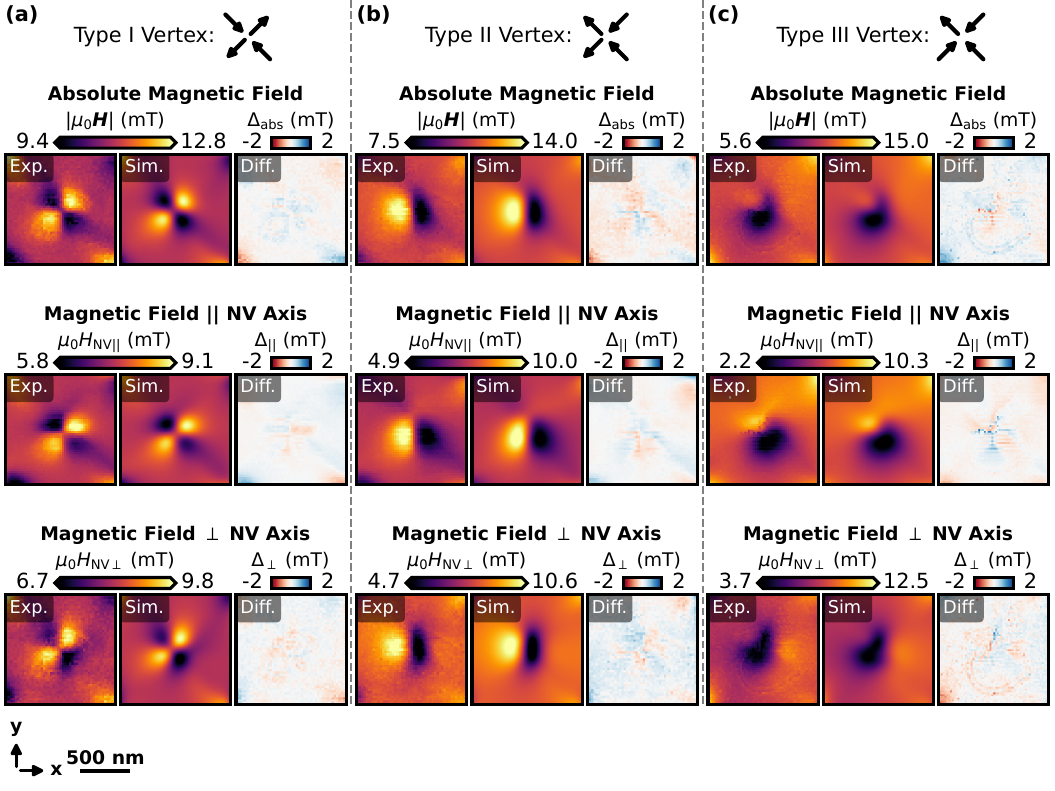}
\caption{
    Comparison of experimental data and final micromagnetic modeling. The modeling used $\mu_0(H_\text{ext,x}, H_\text{ext,y}, H_\text{ext,z}) = (\qty{5.8}{}, \qty{3.7}{}, \qty{-8.7}{})~\qty{}{\milli\tesla}$, $d_\mathrm{MSO} = \qty{65}{\nano\meter}$, $M_\mathrm{s,eff} = \qty{586}{\kilo\ampere\per\meter}$, and the NV angles $(\varphi, \theta) = (\qty{277}{\degree}, \qty{54.75}{\degree})$. Data is shown for the vertices marked in Fig.~\ref{fig:overview} as type I (a), type II (b), and type III (c). The orientation of the nanomagnets' magnetization is indicated by arrows at the top of each respective subfigure.
    The first row displays the magnetic field magnitude $|\mu_0\boldsymbol{H}|$. The second row shows the parallel component $\mu_0 H_{\text{NV}\parallel}$, and the third row presents the perpendicular component $\mu_0 H_{\text{NV}\perp}$, all relative to the NV axis.
    In each subfigure, the first column contains the experimental data (Exp.), the second column shows the corresponding simulation (Sim.), and the third column illustrates the deviation \(\Delta\) (\text{Diff.} = \text{Exp.} - \text{Sim.}).}
    \label{fig:details}
\end{figure*}

Figure~\ref{fig:overview} displays overview scans covering an area of approximately $\qtyproduct{10 x 10}{\micro\meter}$ in the central regions of the $\qtyproduct{25 x 100}{\micro\meter}$ ASI patterns \textit{S-A} (top row) and \textit{S-B} (bottom row). These scans are performed following  magnetic field pulses, which we define as the application of an external field that is subsequently reduced to a low remnant field to perform the measurement. These field pulses are applied along the x-direction for a few seconds, with a magnitude near the respective coercive fields. Prior to each field pulse, the structures were saturated using a strong external field in the x-direction ($\geq \qty{+60}{\milli\tesla}$).

\begin{figure*}[htb]
    \centering
    \includegraphics{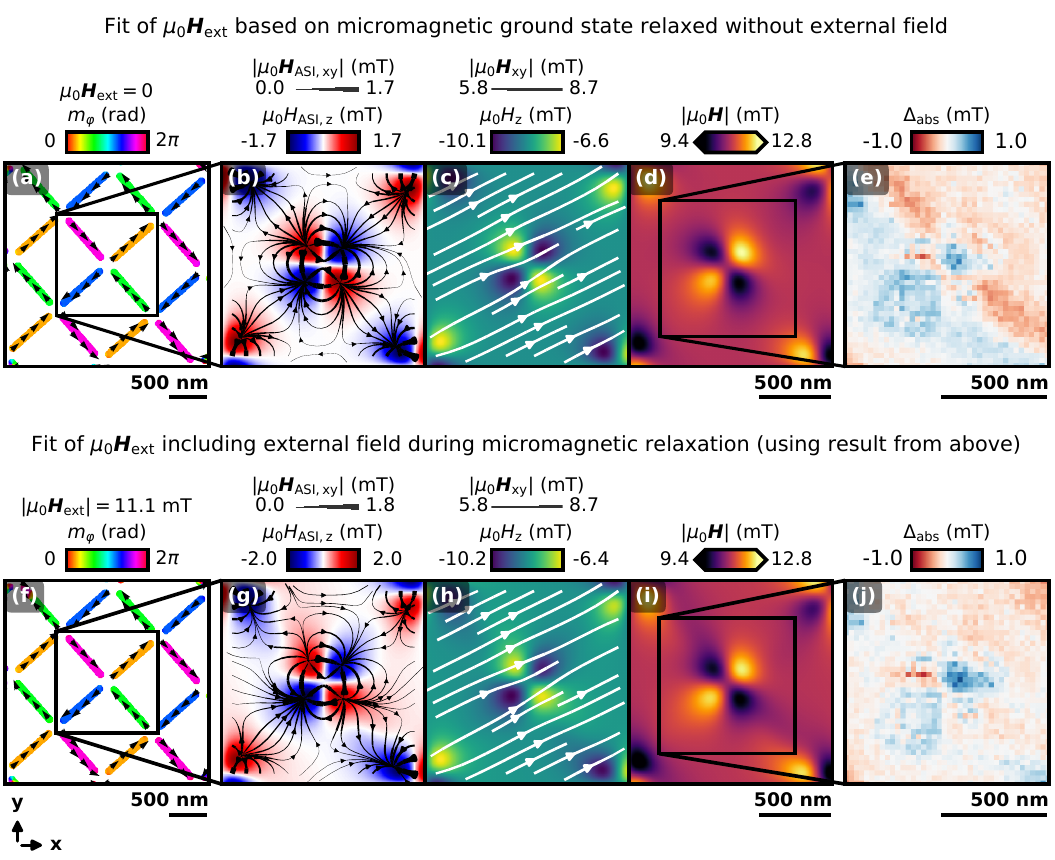}
\caption{
    Illustration of the iterative procedure to determine the external magnetic field $\mu_0\boldsymbol{H}_\text{ext}$ and the corresponding magnetic state, exemplarily shown for the type I vertex highlighted in Fig.~\ref{fig:overview}.
    First Row ((a) - (e)): Fit of $\mu_0\boldsymbol{H}_\text{ext}$ assuming zero external field during micromagnetic relaxation.
    (a) Micromagnetic simulation of the relaxed ground state at $\mu_0\boldsymbol{H}_\text{ext} = 0$, represented by the azimuthal angle $m_\varphi$ of $\boldsymbol{m}$.
    (b) Stray field $\mu_0\boldsymbol{H}_\text{ASI}$, computed \qty{178}{\nano\meter} above the surface of the magnetic layer, in the indicated region.
    (c) Total magnetic field $\mu_0\boldsymbol{H} = \mu_0\boldsymbol{H}_\text{ASI} + \mu_0\boldsymbol{H}_\text{ext}$, fitted to experimental data by minimizing the error $\Delta_{||}^2 + \Delta_\perp^2$.
    (d) Absolute magnetic field $|\mu_0\boldsymbol{H}|$ derived from the fitted vector field in panel (c).
    (e) Deviation $\Delta_\text{abs}$ between (d) and the experimental absolute field data. A systematic error is evident, attributed to neglecting the influence of the external field on the magnetization during relaxation.
    Second Row ((f) - (j)): Fit of $\mu_0\boldsymbol{H}_\text{ext}$ including external field during micromagnetic relaxation (using results from the first iteration).
    (f) Micromagnetic simulation incorporating the fitted $\mu_0\boldsymbol{H}_\text{ext}$ from the first iteration during relaxation.
    (g) Corresponding stray field $\mu_0\boldsymbol{H}_\text{ASI}$, which now exhibits asymmetry induced by the external field.
    (h) Total field $\mu_0\boldsymbol{H}$.
    (i) Absolute field $|\mu_0\boldsymbol{H}|$, showing improved agreement with the experiment (compare to Fig.~\ref{fig:details}(a)).
    (j) Deviation $\Delta_\text{abs}$, showing reduced systematic error.
    }
    \label{fig:iteration}
\end{figure*}

For each sample, the first panel shows a quenching scan after applying a field pulse just below its coercive field (\qty{-35.8}{\milli\tesla} for \textit{S-A} and \qty{-33.9}{\milli\tesla} for \textit{S-B}). During scanning, the stray field in the vertex region vectorially adds to the applied external field, either enhancing or reducing the off-axis component depending on the local nanomagnet orientation. The bias field for these measurements has a negative z-component of approximately \qty{-8.2}{\milli\tesla}, a value determined by subsequent micromagnetic modeling (see Fig.~\ref{fig:iteration}). This leads to a higher fluorescence count rate (i.e., reduced quenching) at magnetic north poles compared to south poles. Individual vertices are framed, and their types are indicated. While \textit{S-A} shows a fully polarized type II lattice, parts of \textit{S-B} have already started to transition to a type I configuration.

Panels (b) and (f) depict the corresponding opposite states after field pulses slightly above the coercive fields (\qty{-40.3}{\milli\tesla} for \textit{S-A}, \qty{-38.5}{\milli\tesla} for \textit{S-B}). Again, \textit{S-A} retains a uniform type II lattice, whereas \textit{S-B} exhibits localized type I regions. The narrower field range required for switching in \textit{S-A} suggests weaker inter-element coupling, which allows individual nanomagnets to reverse more independently.

Panels (c)–(d) and (g)–(h) present quenching and lifted cODMR scans of the same states after applying field pulses approximately equal to the coercive fields (\qty{-38.3}{\milli\tesla} for \textit{S-A}, \qty{-36.3}{\milli\tesla} for \textit{S-B}). To maintain sufficient ODMR contrast for ESR frequency extraction, the cODMR scans are recorded at a lift height of \qty{100}{\nano\meter}, thereby reducing the off-axis field strength. In both samples, type I vertices dominate, separated by chains of type II vertices that delineate local lattice boundaries - more clearly visible in panel (h). Violations of the ice rule (here exclusively type III vertices) emerge, typically located at the endpoints of type II chains (also referred to as Dirac strings \cite{mengotti_real-space_2011}). These energetically costly violations underscore the system’s preference for closed loops, reflecting the topological constraints inherent to artificial spin ice systems.
Comparing the two samples, \textit{S-A} exhibits a significantly higher number of ice-rule violations than \textit{S-B} (9 vs. 1, within the scanned region), consistent with the weaker coupling strength between its nanomagnets.

The formation of horizontal type-II chains separating domain regions of ground-state type-I vertices under a single field pulse (Fig.~\ref{fig:overview}(g,h)) is further investigated via micromagnetic simulations (see Section~S8 in the SM for details). While field-driven vertex transitions in square ASI are well known from continuous field-cycling or rotating-field protocols \cite{wang_artificial_2006, PhysRevLett.98.217203, PhysRevLett.101.037205, PhysRevLett.105.017201, Budrikis_2012}, our simulations confirm that a single, interrupted reverse pulse is sufficient to trigger line-by-line conversions from type~II to type~I vertices. 

We utilize Mumax3 \cite{vansteenkiste_design_2014} within the software platform Aithericon \cite{aithericon} to simulate the stray field generated by the micromagnetic ground states of individual vertices in sample \textit{S-A}. This simulation is then compared with the cODMR scan data. Further details on the micromagnetic simulations are provided in Section S3 of the SM.

For high-resolution cODMR scans and quantitative modeling, three distinct vertices (of types I, II, and III), highlighted in gold in Fig.~\ref{fig:overview}(c,d), were selected. The modeling employs an iterative optimization procedure that minimizes the mean combined deviation $\Delta_{\parallel}^2 + \Delta_\perp^2$. Here, $\Delta_{\parallel}$ and $\Delta_\perp$ represent the differences between the simulated and measured magnetic field components $\mu_0H_{\text{NV}\parallel}$ and $\mu_0H_{\text{NV}\perp}$, respectively. A comprehensive graphical overview of the entire modeling algorithm is provided in Section S4 of the SM.

The parameters optimized during this process include the external bias field $\boldsymbol{H}_\text{ext}$, the NV-to-sample distance, the NV axis orientation and the effective saturation magnetization $M_\mathrm{s,eff}$. This $M_\mathrm{s,eff}$ accounts for both intrinsic material and geometrical properties as well as geometrical imperfections of the nanomagnets.

Figure~\ref{fig:iteration} illustrates the first step of this modeling, exemplified by the type I vertex. In this initial step, $\boldsymbol{H}_\text{ext}$ and the corresponding magnetic ground state, here depicted using the normalized magnetization $\boldsymbol{m}$, are determined in two iterations. Initially, the magnetization is simulated for $\mu_0\boldsymbol{H}_\text{ext} = 0$ to obtain the stray field $\boldsymbol{H}_\text{ASI}$ at an assumed measurement height of $d_\mathrm{NV-Py} \approx \qty{178}{\nano\meter}$ (panels (a)-(b)). This estimate combines an approximation of the magnetic stand-off distance in contact $d_\mathrm{MSO} \approx \qty{75}{\nano\meter}$ \cite{tetienne_nature_2015, hedrich_parabolic_2020, xu_minimizing_2025}, a lift height of $d_\mathrm{lift} = \qty{100}{\nano\meter}$, and a capping layer thickness $d_\mathrm{Al} = \qty{3}{\nano\meter}$.

Subsequently, $\Delta_{||}^2 + \Delta_\perp^2$ is minimized by adding $\boldsymbol{H}_\text{ext}$ and multiplying $\boldsymbol{H}_\text{ASI}$ by a factor $c$, which yields the effective saturation magnetization $M_\mathrm{s, eff} = c \cdot M_\mathrm{s}$. It is important to note that $M_\mathrm{s,eff}$ at this stage is sensitive to the assumed $d_\mathrm{MSO}$ and is therefore preliminary; a final value is determined after the true $d_\mathrm{MSO}$ has been considered in the subsequent process. Additionally, the spatial alignment between the simulated and measured data is optimized automatically by numerically adjusting lateral ($x$ and $y$) shifts of the simulation map and interpolating it to match the measurement coordinates, as part of the minimization algorithm.
To minimize the influence of potential drift during the measurement, the optimization is confined to a \qtyproduct{700x700}{\nano\meter} region around the vertex.
The resulting magnetic field distribution is presented as a streamline plot in panel (c) and as absolute field magnitude in panel (d). Panel (e) shows the deviation between the simulation and measurement by subtracting simulated from experimental 2D profiles, which reveals a systematic error attributed to neglecting the effect of $\boldsymbol{H}_\text{ext}$ on the magnetization. The mean absolute deviation in Fig.~\ref{fig:iteration}(e) is \qty{0.162}{\milli\tesla}.

The lower row (panels (f)–(j)) repeats the procedure, but now incorporates the external field $\mu_0(H_\text{ext,x}, H_\text{ext,y}, H_\text{ext,z}) = (\qty{6.5}{}, \qty{3.2}{}, \qty{-8.4}{})~\qty{}{\milli\tesla}$ from the first optimization pass into the micromagnetic simulation. This step induces slight asymmetry in $\boldsymbol{H}_\text{ASI}$ (compare panels (b) and (g)), caused by small tilts in the magnetization, particularly at the rounded nanomagnet ends (see Section S9 in the SM). The improved agreement is evident as the systematic deviation in panel (j) is reduced; the mean absolute deviation, for instance, is reduced by approximately \qty{27}{\percent} to \qty{0.119}{\milli\tesla} compared to panel (e). This second iteration yields a slightly refined external field of $\mu_0(H_\text{ext,x}, H_\text{ext,y}, H_\text{ext,z}) = (\qty{6.6}{}, \qty{3.3}{}, \qty{-8.2}{})~\qty{}{\milli\tesla}$. Remaining systematic error at this stage can mainly be attributed to an under- or overestimation of $d_\mathrm{MSO}$ and a small misalignment of the NV orientation.

Subsequently, the actual values of $d_\mathrm{MSO}$ and $M_\mathrm{s,eff}$ are determined by minimizing $\Delta_{\parallel}^2 + \Delta_\perp^2$ as a function of $d_\mathrm{MSO}$. 
To achieve this, the fitting procedure (as performed in the second iteration of Fig.~\ref{fig:iteration}), using the micromagnetic ground state depicted in Fig.~\ref{fig:iteration}(f), is repeated for different fixed values of $d_\mathrm{MSO}$, using a step size of \qty{5}{\nano\meter}. For each $d_\mathrm{MSO}$ value, the minimum deviation $\Delta_{\parallel}^2 + \Delta_\perp^2$ achieved by this fitting procedure is recorded.
The best agreement with the measured data is found for $d_\mathrm{MSO} = \qty{65}{\nano\meter} \pm \qty{5}{\nano\meter}$ (see Section S5 in the SM for further details).

In the final stage, the uncertainty of the NV orientation is acknowledged by finding the minimum fitting loss $\Delta_{\parallel}^2 + \Delta_\perp^2$ across a grid of NV orientation angles $\varphi$ and $\theta$ spanning $\pm\qty{10}{\degree}$ around their respective nominal values, thereby performing an in-situ calibration of the NV orientation. We use the same fitting procedure as in the previous step, which allows us to track the effect of NV misalignment on the fitting parameters and evaluate them at the optimal orientation. The loss function exhibits a distinct minimum at $(\varphi, \theta) = (\qty{277}{\degree} \pm \qty{1}{\degree}, \qty{54.75}{\degree} \pm \qty{1}{\degree})$, indicating a minor misalignment of $\Delta\varphi = \qty{7}{\degree}$ relative to the nominal crystallographic direction ($\varphi = \qty{270}{\degree}$). At this calibrated NV orientation, we extract the final fitting parameters, yielding an external magnetic field of $\mu_0(H_\text{ext,x}, H_\text{ext,y}, H_\text{ext,z}) = (5.8, 3.7, -8.7)~\qty{}{\milli\tesla}$ and an effective saturation magnetization of $M_\mathrm{s,eff} = \qty{586}{\kilo\ampere\per\meter}$. Regarding the orientation-dependent variation in the fitted external magnetic field, we find maximum local gradients ranging between \qtyrange{0.13}{0.16}{\milli\tesla\per\degree\squared}, depending on the field component (see Section~S6 of the SM for details). 

Finally, the error in the extracted saturation magnetization is primarily governed by its dependence on $d_\mathrm{MSO}$. Using the local slope $\frac{\partial M_\mathrm{s,eff}}{\partial d_\mathrm{MSO}} = \qty{4.9}{\kilo\ampere\per\meter\per\nano\meter}$ extracted from Fig. S5 of the SM, we estimate an error bound of $\pm\qty{25}{\kilo\ampere\per\meter}$ based on a grid uncertainty of $\pm\qty{5}{\nano\meter}$ in $d_\mathrm{MSO}$. Additionally, accounting for stochastic fluctuations of $\pm 5\%$ determined via a line cut through a type II chain from Fig.~\ref{fig:overview}(d) (see Section S7 in the SM) adds a standard contribution of $\qty{586}{\kilo\ampere\per\meter} \cdot \qty{\pm5}{\percent} \cdot \frac{1}{\sqrt{3}} \approx \pm\qty{17}{\kilo\ampere\per\meter}$, leading to an overall error bound of $\pm\qty{42}{\kilo\ampere\per\meter}$.

Notably, the obtained value of $M_\mathrm{s,eff} = \qty{586}{\kilo\ampere\per\meter} \pm \qty{42}{\kilo\ampere\per\meter} $ is significantly lower than literature values for Ni\textsubscript{81}Fe\textsubscript{19}, such as $M_\mathrm{s}^\mathrm{lit} = \qty{860}{\kilo\ampere\per\meter}$ \cite{nahrwold_structural_2010}. This discrepancy suggests a deviation between the idealized geometry used in simulations and the actual fabricated structures, likely due to features such as a conical cross-section, which reduces the magnetic volume. An AFM-derived cross-section (see Section S1 in the SM) further supports this interpretation.

Figure~\ref{fig:details} provides a comprehensive comparison of measured magnetic field components and final micromagnetic modeling for three distinct vertex types (I, II, and III, highlighted in gold in Fig.~\ref{fig:overview}). Columns (a-c) depict each vertex, with rows displaying the absolute magnetic field ($|\mu_0\boldsymbol{H}|$), the parallel component ($\mu_0 H_{\text{NV}\parallel}$), and the perpendicular component ($\mu_0 H_{\text{NV}\perp}$) relative to the NV axis. Each panel shows experimental data (Exp.), corresponding simulation (Sim.), and their difference (Diff. = Exp. - Sim.).

Excellent agreement is observed between experimental maps and simulations across all vertex types and field components. This confirms the accuracy of magnetic field determination via scanning NV magnetometry and the high fidelity of our modeling and iterative optimization of $\boldsymbol{H}_\mathrm{ext}$, $M_\mathrm{s,eff}$, and $d_\mathrm{MSO}$, as well as the NV angles $\varphi$ and $\theta$.
The difference maps ($\Delta$) show minimal residual errors, typically within \qty{\pm 1}{\milli\tesla}. Remaining discrepancies are likely due to minor unmodeled structural imperfections, mechanical drift during the experimental scans, and the approximation that the experimental data is obtained in a plane while the probe in reality follows the sample's topography.

In summary, we demonstrated that scanning single-NV magnetometry is a powerful and non-invasive technique for resolving local dipolar coupling fields in artificial spin ice systems with nanoscale precision. Fluorescence quenching imaging, with contrast tunable via an external magnetic field, provides a fast and efficient method for identifying the Ising-like states of individual nanomagnets. This allows for rapid statistical investigations such as quantifying ice-rule violations across extended regions and linking them to geometric lattice parameters. Continuous-wave ODMR offers access to both axial and transverse magnetic field components at the nanoscale, with sensitivities reaching down to a few \qty{}{\micro\tesla\per\sqrt{\hertz}}. Our approach enables direct, quantitative comparison with micromagnetic simulations, supporting precise determination of the local bias field, identification of small magnetization tilts induced by external perturbations, the extraction of the effective saturation magnetization and an in-situ calibration of the NV orientation angles via an iterative fitting procedure. This quantitative capability holds significant implications for advancing the design of magnetic nanodevices. By enabling detailed characterization of dipolar interactions, including effects from structural imperfections and material properties arising from nanostructuring, our work paves the way for novel spintronic and magnonic devices where precisely tailored stray fields are crucial for functionality, potentially facilitating interactions with other quantum or classical systems \cite{Sultana_2025}.
\\
\section*{Conflict of Interest}
The authors have no conflicts to disclose.
\\
\section*{Acknowledgements}
Work at the University of Delaware was supported by the National Science Foundation under Grant No. 2339475. Work at RPTU was supported by the Deutsche Forschungsgemeinschaft (DFG, German Research Foundation) within the Transregional Collaborative Research Center TRR 173/3–268565370 ”Spin +X” (Projects A12 and B15). RPTU acknowledges the use of our Scanning NV Magnetometer, funded by the Deutsche Forschungsgemeinschaft (DFG, German Research Foundation) - 491229782 within the major instrumentation initiative “Spin-based quantum light microscopy (SQLM)”.
\\
\section*{Data Availability}
The data that support the findings of this study are openly available at \href{https://doi.org/10.17605/OSF.IO/M4TGH}{10.17605/OSF.IO/M4TGH}.

\providecommand{\noopsort}[1]{}\providecommand{\singleletter}[1]{#1}%

\end{document}